\documentclass{article}

\usepackage{amssymb}

\usepackage{amsmath}
\usepackage{graphics}
\usepackage{graphicx}

\newcommand{\bea}{\begin{eqnarray}}

\newcommand{\eea}{\end{eqnarray}}

\newcommand{\be}{\begin{equation}}

\newcommand{\ee}{\end{equation}}

\newcommand{\rt}[1]{{}}

\newlength{\szovszel}
\newlength{\slashszel} 
\newcommand*{\sls}[1]{\mbox{%
    \settowidth{\szovszel}{\ensuremath{#1}}%
    \settowidth{\slashszel}{\ensuremath{\slash}}%
    \hspace*{0.5\szovszel}%
    \hspace*{-0.5\slashszel}%
    \slash%
    \hspace*{-0.5\szovszel}%
    \hspace*{-0.5\slashszel}%
    \ensuremath{#1}%
  }}

\begin{document}

\title{Renormalisation Group determination\\ of scalar mass bounds in a simple Yukawa-model}

\author{A. Jakov\'ac, I. Kaposv\'ari and A. Patk\'os\footnote{Presented at the Gribov-85 Conference, Chernogolovka, Russia, June 24-28, 2014, to appear in the Gribov Memorial Volume, eds. Yu. Dokshitzer, P. L\'evai and J. Nyiri}\\
Institute of Physics, E\"otv\"os University\\
H-1117, P\'azm\'any P\'eter s\'et\'any 1/A, Budapest, Hungary}
\vfill
\maketitle
\begin{abstract}
The scalar mass is determined in the simplest scalar-fermion Yukawa-model in the whole range of stability of the scalar potential. Two versions of the Functional Renormalisation Group (FRG) equations are solved, where also composite fermionic background is taken into account. The close agreement of the results with previous studies taking into account  exclusively the effect of the scalar condensate, supports a rather small systematic truncation error of FRG due to the omission of higher dimensional operators.  
\end{abstract}

\section{Motivation: from "Moscow-zero" to Higgs-mass bounds}

The discovery of the "Moscow-zero" \cite{abrikosov54} (alias Landau-pole) appears from modern perspective as a true Archimedean "fixed point" on the historical path of renormalized quantum field theories. Its real importance has been recognized after desperate discussions concerning the very sense of quantum field theory. The present conference gives the opportunity to quote from the history notes of A.A. Anselm \cite{anselm-www} that researchers of the Theory Division of the Leningrad (at present Petersburg) Nuclear Physics Institute  under the "natural scientific leadership" of V.N. Gribov contributed "many important papers during the 50's and the beginning of the 60's to the development of the "null-charge" problem." And the interest of Volodia Gribov did not diminish with time \cite{gribov94}.

At present two alternative attitudes are taken concerning the exact nature of the field theories which in the framework of (resummed) perturbation theory appear to possess Landau-pole. The concept of asymptotic safety \cite{weinberg79b}‪
envisages the existence of an UV-attractive fixed point in the renormalisation flow of the model under study. The energy scale of this fixed point should be lower than the scale of the Landau-pole. Assuming the existence of such a fixed point in the joint theory of gravity and the Standard Model at the scale of the Planck-mass a quite accurate prediction(!) of the mass of the Higgs particle was achieved \cite{shaposhnikov10}. Some recent papers \cite{litim14} express very optimistic views concerning the existence of a UV-safe completion of the Standard Model on its own, but to present date no such interactive fixed point could have been found. 

The other alternative phrased in form of the triviality conjecture \cite{luescher87} accepts the strict existence of a Landau-type singularity in the scalar sector of the Standard Model. This view gains support in case of the four-dimensional $\Phi^4$ theory from non-perturbative numerical simulations. In order to avoid the manifestation of the effects of the  singularity one has to restrict the momentum range of the modes participating in the dynamics of the model by applying some sort of momentum cut-off chosen below the location of the Landau-singularity.
The scale of the cutoff indicates where the presently known fields and  interactions are expected to reach the edge of validity. The Lagrangean of these effective models \cite{weinberg79a} is allowed to contain also operators which are perturbatively (in the weak coupling regime) non-renormalizable. Their coupling strengths are proportional to some negative powers of the cut-off. The detailed choice of the Lagrangean at the cut-off exerts influence on the renormalisation flow towards the infrared. Conversely, the solution of the model which is designed to a certain predetermined set of low energy observables can be extended only up to a certain maximal momentum scale. In this sense the value $\Lambda$ of the maximal momentum is part of the physical data set characterizing the effective model. This approach has been used extensively for deriving upper bounds for the Higgs-particle in the scalar sector of the Standard Model \cite{maiani78,dashen83,lindner86,kuti88,hambye97}. 

After the discovery of the Higgs-particle with mass 125GeV the actual question one asks is somewhat reoriented. Now the lower bound arising from requiring the stability of the Higgs potential is more in the focus. In particular, the possibility of the higher dimensional operators playing an essential role in the stability of the model is now actively investigated \cite{kuti07,gies14,gies15a,gies15b,chu15}. In finding a reliable answer the accuracy of the renormalisation group equations (RGE) connecting the low energy momentum range with the operator content of the theory defined at the cutoff is of outstanding importance. Since all methods of RGE-construction involve some approximations (truncations) it is of great interest to have a clear estimate on the sources and sizes of the truncation errors. 

At present the most widely used form of the RGE is the Wetterich-Morris equation    \cite{wetterich91,wetterich93-3,morris94-3}, which decribes the rate of evolution of the effective quantum action $\Gamma_k$ at scale $k,~(t=\ln k)$:
\be
\partial_t\Gamma_k[\phi]=\frac{1}{2}{\textrm{Str}}\left\{(\partial_tR_k)\left[\Gamma^{(2)}_k+R_k\right]^{-1}\right\}.
\label{wetterich-morris}
\ee
The operation $\hat\partial_k$ appearing in front of the expressions on the right hand side of (\ref{LPA-RGE}) requires the computation of the $k$-derivatives of the whole expression behind it, but taking into account only the $k$-dependence of the regulator functions $R_k^F(q)$ and $R_k^B(q)$.
The functional super-trace on the right hand side contains the second functional derivative of $\Gamma_k[\Phi]$ depending on a set of fields denoted by $\Phi$, together with an appropriately chosen infrared regulator function $R_k$. The super-trace over fermionic loops involves an extra minus sign relative to the bosonic loops. 
The aim of the present contribution is to construct and solve the RGE of a simple scalar-Yukawa model in an improved approximation scheme to this equation which goes beyond the accuracy of the previous investigations. in particular Ref.\cite{gies14}, where approximate solutions of Eq.(\ref{wetterich-morris}) were constructed with a more restricted background configuration. The bounds found for the scalar mass upon fixed values for the fermion mass and the scalar condensate will be compared to the results of Ref.\cite{gies14}.

\section{The scalar-fermion Yukawa model and its RGE}

The solution of (\ref{wetterich-morris}) is attempted for the scalar-fermion Yukawa model in Euclidean space-time using the following effective action Ansatz defining the field theory at scale $k$:
\be
\Gamma_k=\int_x\left[Z_{\psi k}\bar\psi\sls\partial\psi+
\frac{1}{2}Z_{\sigma k}(\partial_m\sigma)^2+h_k\sigma\bar\psi\psi+U_k(\rho)\right],\qquad \rho=\frac{1}{2}\sigma^2.
\label{action-ansatz}
\ee
A non-zero condensate $v_0=Z_{\sigma 0}^{1/2}<\sigma>=\sqrt{2Z_{\sigma 0}\rho_0}$ in the scalar field $\sigma$ violates the symmetry of the action under the discrete $\gamma_5$ transformation of the fermionic field $\psi$:
\be
\sigma(x)\rightarrow -\sigma(x),\qquad \psi(x)\rightarrow \gamma_5\psi(x),\quad \bar\psi(x) \rightarrow -\bar\psi(x)\gamma_5.
\label{gamma5-symmetry}
\ee
The condensate generates a mass for the fermi-field: $m_\psi=h_0v_0$. The index '0' in 
$h, \rho_0, Z_{\sigma 0}$ and $v_0$ emphasizes that the values to be used for the computation of these quantities should be evaluated from the action $\Gamma_{k=0}$, where $\Gamma_{k=\Lambda}$ arrives when running with help of (\ref{action-ansatz}) from the cutoff $\Lambda$ to $k=0$. In order to maintain maximal analogy between our model and the top-Higgs sector of the Standard Model, in the discussion of the solution below we shall fix the $k=0$ values of the vacuum condensate to $v_0=246$GeV and the fermion mass to $m_\psi=173$GeV. In this preliminary report we shall neglect the running of the field renormalisations and set $Z_{\phi k}=Z_{\psi k}=1$, which is the so-called Local Potential Approximation (LPA) to (\ref{wetterich-morris}).

The second functional derivative of the effective action is represented as a 3x3 matrix:
\be
\begin{pmatrix}
{\Gamma^{(2)}_{\sigma\sigma}} & \Gamma^{(2)}_{\sigma\psi}&\Gamma^{(2)}_{\sigma\bar\psi^T}\\
\Gamma^{(2)}_{\psi^T\sigma}&{\Gamma^{(2)}_{\psi^T\psi}}&{\Gamma^{(2)}_{\psi^T\bar\psi^T}}\\
\Gamma^{(2)}_{\bar\psi\sigma}&{\Gamma^{(2)}_{\bar\psi\psi}}&{\Gamma^{(2)}_{\bar\psi\bar\psi^T}}
\end{pmatrix},
\ee
where for the representation of the fields we use the column-vector $(\sigma(x), \psi(x), \bar\psi^T(x))$ which corresponds to the Gor'kov-Nambu representation of the fermi-field. As a first operation one can transform this matrix into a separate block-diagonal form in the scalar and the fermion-sector, which leads for the super-trace to
\bea
&&\frac{1}{2}{\textrm{Str}}\log(\Gamma^{(2)}+R_k)=
-\frac{1}{2}{\textrm{Tr}}\log(\Gamma^{(2)}_{\Psi^T\Psi}+R_k^F)
+\frac{1}{2}{\textrm{Tr}}\log(\Gamma^{(2)}_{\sigma\sigma}-\Gamma^{(2)}_{\sigma\Psi}\Gamma^{(2)-1}_{\Psi^T\Psi}\Gamma^{(2)}_{\Psi^T\sigma})\nonumber\\
&&
~~~~~~~~~~~~~~~~~~~~~~~~~~~=-\frac{1}{2}{\textrm{Tr}}\log(\Gamma^{(2)}_{\Psi^T\Psi}+R_k^F)+\frac{1}{2}{\textrm{Tr}}\log(\Gamma^{(2)}_{\sigma\sigma}+R_k^B)\nonumber\\
&&
~~~~~~~~~~~~~~~~~~~~~~~~~~~~~~~+\frac{1}{2}{\textrm{Tr}}\log\left[1-\left(\Gamma^{(2)}_{\sigma\sigma}+R_k^B\right)^{-1}\Gamma^{(2)}_{\sigma\Psi}
\left(\Gamma^{(2)}_{\Psi^T\Psi}+R_k^F\right)^{-1}
\Gamma^{(2)}_{\Psi^T\sigma}\right].
\label{super-tracelog-1}
\eea
Here $\Gamma^{(2)}_{\Psi^T\Psi}$ is a compact notation for a hipermatrix in the doubled bispinor field basis.

\begin{figure} 
\includegraphics[width=3.5cm]{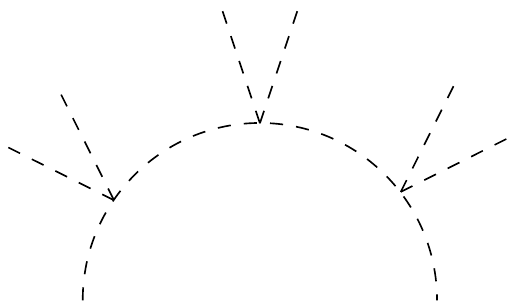}
\includegraphics[width=3.5cm]{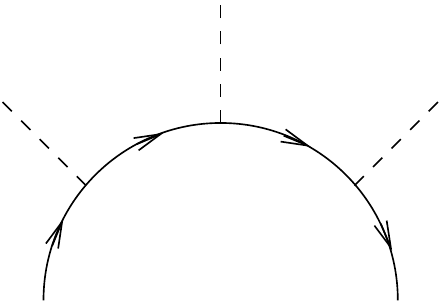}
\includegraphics[width=3.5cm]{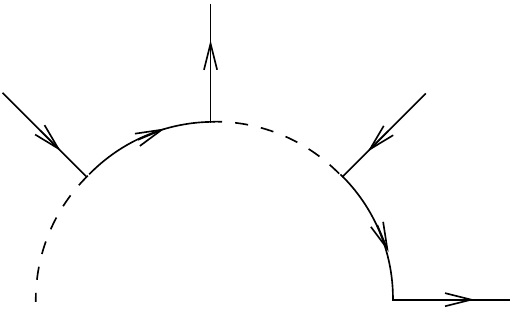}
\caption{Propagator sequences building the one-loop contributions to the tracelog of (\ref{super-tracelog-1}). On the left a piece of a scalar loop (dashed curves) immersed in a scalar condensate, in the middle a fermion loop (continuous curve) immersed in a scalar condensate, on the right a loop constructed of alternating sequence of fermion and scalar propagators immersed in a composite fermionic condensate is represented.}
\label{one-loop-contribution-tracelog}
\end{figure}
Each term has a diagrammatic representation. The first two terms correspond to the free fermion and free boson contributions to the super-tracelog computed in the background $v_k$. The last term is the sum of the infinite series of 1-loop contributions containing an increasing number of alternating boson and fermion propagators. At each vertex of fermion-to-boson transformation an external fermion leg joins the loop, due to the assumed presence of a composite fermionic background (see Fig.\ref{one-loop-contribution-tracelog}). 

When one explicitly displays the matrix operations in the bispinor indices in the doubled fermionic sector, one finds for the last term of the previous equation
\bea
&&{\textrm{Tr}}\log\left[1-\left(\Gamma^{(2)}_{\sigma\sigma}+R_k^B\right)^{-1}\Gamma^{(2)}_{\sigma\Psi}
\left(\Gamma^{(2)}_{\Psi^T\Psi}+R_k^F\right)^{-1}
\Gamma^{(2)}_{\Psi^T\sigma}\right]\nonumber\\
&=& 
{\textrm{Tr}}\log\left[1-\left(\Gamma^{(2)}_{\sigma\sigma}+R_k^B\right)^{-1}\left[\Gamma^{(2)}_{\sigma\psi}\left(\Gamma^{(2)}_{\bar\psi\psi}+R_k^F\right)^{-1}\Gamma^{(2)}_{\bar\psi\sigma}+
\Gamma^{(2)}_{\sigma\bar\psi^T}\left(\Gamma^{(2)}_{\psi^T\bar\psi^T}+R_k^F\right)^{-1}\Gamma^{(2)}_{\psi^T\sigma}\right]\right].
\label{eq:regrouping}
\eea

In previous papers the super-trace was evaluated on a background exclusively consisting of a nonzero $\sigma$-condensate \cite{gies14}. On such a background $\Gamma^{(2)}_{\sigma\Psi}=\Gamma^{(2)}_{\Psi^T\sigma}=0$ and the right hand side of the RGE simply reduces to the combination of a pure fermion- and a pure boson-loop. A  more general choice containing both $\rho_k$ and the pointlike fermion condensate $(\bar\psi\psi)_k$, which moreover can be made compatible with the field equation
\be
\frac{\delta\Gamma_k}{\delta\sigma}\bigl|_{\sigma=v_k,\psi=\psi_k}=h_k(\bar\psi\psi)_k+v_kU_k'(\rho_k)=0,
\label{effective-field-equation}
\ee
might be more adequate and anyhow checks the robustness of the lower bounds obtained earlier.
In our recent publications clear arguments were put forward \cite{jakovac13,jakovac15} how one can reconcile the Grassmannian nature of the fermi fields with a nonzero coarse grained
$[(\bar\psi\psi)_k]^n, ~n>1$ background. Using the explicit expressions of the scalar-fermion (Yukawa) vertices and of the fermion propagators one quickly evaluates the contribution of the last term on the right hand side of (\ref{eq:regrouping}) on a space-independent $v_k$ and $(\bar\psi\psi)_k$ background:
\bea
&&
{\textrm{Tr}}\log\left[1-\left(\Gamma^{(2)}_{\sigma\sigma}+R_k^B\right)^{-1}\left(\Gamma^{(2)}_{\sigma\psi}\left(\Gamma^{(2)}_{\bar\psi\psi}+R_k^F\right)^{-1}\Gamma^{(2)}_{\bar\psi\sigma}+
\Gamma^{(2)}_{\sigma\bar\psi^T}\left(\Gamma^{(2)}_{\psi^T\bar\psi^T}+R_k^F\right)^{-1}\Gamma^{(2)}_{\psi^T\sigma}\right)\right]\nonumber\\
&&
~~~~~~~~~~~~~=V_d\frac{1}{2}\int_q\log\left\{1-h_k^2\frac{1}{q_R^2+m_{B0}^2}\frac{2m_\psi(\bar\psi\psi)_k}{q_R^2+m_\psi^2}\right\}.
\eea
Here $V_d$ is the $d$-dimensional quantisation volume, $m_{B0}^2=U'_k(\rho_k)+2\rho_kU''_k(\rho_k), m_\psi^2=2h_k^2\rho_k$ and $q_R^2$ stands for the infrared regularized momentum-dependent part of the inverse propagator: $q^2+R_k(q)$ (one chooses different regulators for bosons and fermions).

Within the LPA Ansatz(\ref{action-ansatz}) one finds the following RGE for the potential energy of the model after combining the three contributions on its right hand side:
\bea
&&\partial_k[U_k(\rho_k)+h_kv_k(\bar\psi\psi)_k]\nonumber\\
&=&\frac{1}{2}\hat\partial_k\int_q\left[-4\log(q_R^2+m_\psi^2)+\log(q_R^2+m_{B0}^2)+\log\left\{1-h_k^2\frac{1}{q_R^2+m_{B0}^2}\frac{2m_\psi(\bar\psi\psi)_k}{q_R^2+m_\psi^2}\right\}\right].
\label{LPA-RGE}
\eea

There are several strategies available how to deal with this equation. We shall study two variants below. The first is simply to substitute (\ref{effective-field-equation}) on the right hand side and arrive there at an expression which depends exclusively on $\rho_k$. This approach will be called below {\it version A} and has the following RGE:
\be
\partial_kh_k=0,
\ee
\be
\partial_kU_k(\rho_k)
=\frac{1}{2}\hat\partial_k\int_q\left\{-5\log(q_R^2+2h_k^2\rho_k)+\log\left[(q_R^2+\tilde U(\rho_k))(q_R^2+2h_k^2\rho_k)+4h_k^2\rho_kU'(\rho_k)\right]\right\}.
\label{LPA-version-1}
\ee
 This step will not influence the change in the other couplings since the change of the equation of state is higher order in an infinitesimal step of scale change. 
The other reasonable attitude is to expand the right hand  side of (\ref{LPA-RGE}) in power series of $(\bar\psi\psi)_k$, equate the coefficients of the first power of the expansion on the two sides and substitute (\ref{effective-field-equation}) only for the higher powers. This leads to the following coupled set of RGE:
\bea
\partial_kU_k(\rho_k)
&=&\frac{1}{2}\hat\partial_k\int_q\left\{-5\log(q_R^2+2h_k^2\rho_k)+\log\left[(q_R^2+\tilde U(\rho_k))(q_R^2+2h^2\rho_k)+4h_k^2\rho_kU'(\rho_k)\right]\right\}\nonumber\\
&-&\frac{1}{2}\hat\partial_k\int_q\frac{2h_k^3}{(q_R^2+m_\psi^2)(q_R^2+m_{B0}^2)}\times\frac{2\rho_kU'_k(\rho_k)}{h_k},\nonumber\\
\partial_k(h_kI_k)&=&-\frac{1}{2}\hat\partial_k\int_q\frac{2h_k^3I_k}{(q_R^2+m_\psi^2)(q_R^2+m_{B0}^2)}, \qquad I_k=v_k(\bar\psi\psi)_k.
\label{LPA-version-2}
\eea
 It is obvious that there is an infinite number of further possible approximations to the right hand side of the RGE which are linear in $I_k$. Note that the equation of the Yukawa-coupling one can extract directly from the potential energy density, since one has introduced also a composite fermion background in addition to $\rho_k$.  This way of extracting should be equivalent to the projection with help of the appropriate third functional derivative of $\Gamma_k$ \cite{gies14}.

In the actual computations the so-called Litim regulators were used \cite{litim01}, which allow a quick evaluation of the integrals on the right hand side of RGE. The equations are rewritten next in terms of dimensionless quantities using the index 'r' for them:
\bea
&&v_k=v_rk^{d/2-1},\qquad \psi_k=\psi_rk^{(d-1)/2},\qquad I_k=I_rK^{3d/2-1},\nonumber\\
&&
h_k=h_rk^{2-d/2},\qquad U_k(\rho_k)=k^du_r(\rho_r=\rho_k k^{-d+2}).
\eea
One arrives at a single equation in {\it version A}:
\be
\partial_tu_r+du_r+(2-d)\rho_ru'_r=v_d\left(-\frac{5}{1+\mu_\psi^2}+\frac{2+\mu_\psi^2+\mu_\sigma^2}{(1+\mu_\psi^2)(1+\mu_\sigma^2)+4h_r^2\rho_ru'_r}\right),\qquad h_k=h_\Lambda,
\label{rescaled-u-A}
\ee
and a set of two coupled equations in {\it version B}:
\bea
\partial_tu_r+du_r+(2-d)\rho_ru'_r&=&v_d\Biggl(-\frac{5}{1+\mu_\psi^2}+\frac{2+\mu_\psi^2+\mu_\sigma^2}{(1+\mu_\psi^2)(1+\mu_\sigma^2)+4h_r^2\rho_ru'_r}\nonumber\\
&+&4h_r^2\rho_ru'_r\frac{2+\mu_\psi^2+\mu_\sigma^2}{(1+\mu_\psi^2)^2(1+\mu_\sigma^2)^2}\Biggr).
\label{rescaled-urunning-B}
\eea
and
\be
\partial_th_r^2+(4-d)h_r^2=4h_r^4v_d\frac{2+\mu_\psi^2+\mu_\sigma^2}{(1+\mu_\psi^2)^2(1+\mu_\sigma^2)^2}.
\label{rescaled-hrunning-B}
\ee
In these equations the following notations were introduced:
\be
m_\sigma^2=k^2\mu_\sigma^2,\qquad m_\psi^2=k^2\mu_\psi^2, \qquad v_d=\frac{S_d}{d(2\pi)^d},
\ee
($S_d$ is the surface of the $d$-dimensional unit sphere).

\section{Estimating $m_\sigma(\lambda, \Lambda)$ with help of the solution of RGE}

Whatever parametrisation is used at $k=\Lambda$ the couplings characterizing (\ref{action-ansatz}) should be tuned to arrive at $k=0$ in the broken symmetry phase where the minimum of the potential $U_0$ is taken at $\rho_{0,min}=(246 GeV)^2/2$ and the Yukawa coupling hits the value $h_0=173/246$. Then one determines the second derivative of the potential at the minimum, which determines the value of the scalar mass:
\be
m_\sigma^2=2\rho_{0,min}U''_0(\rho_{0,min}).
\ee
The minimal Ansatz for the potential which is flexible enough for the required tuning of the parameters at $k=\Lambda$ is a quartic potential. It has different parametrisations in the symmetric (SYM) and the broken symmetry (SB) regimes:
\be
u^{SYM}_r(\rho_r)=\lambda_{1k}\rho_r+\frac{1}{2}\lambda_{2k}\rho_r^2,
\qquad 
u_r^{SB}(\rho_r)=\frac{1}{2}\lambda_{2k}(\rho_r-\kappa_k)^2.
\ee
The stability of the potential requires in both phases $\lambda_{2k}\geq 0$. In the symmetric phase $\lambda_{1k}>0$. In the broken symmetry phase one has
\be
\lim_{k\rightarrow 0}2k^2\kappa_k\Lambda^2=v_{0,min}^2, \qquad m_\sigma^2=\lambda_{2,0}\Lambda^2.
\ee

In {\it version A} one chooses the fixed value $h_0=h_k=h_\Lambda$ and by tuning $\lambda_{1,\Lambda}$ and subsequently $\kappa_\Lambda$ one arrives at $k=0$ to the physical value of the condensate. When $\lambda_{2,\Lambda}$ is not too large, one crosses over from the symmetric to the symmetry broken phase at some intermediate scale $\lambda_{1,crit}=0$ and one has to continue with the symmetry breaking parametrisation of the potential. For large enough $\lambda_{2,\Lambda}$ one starts straight in the SB phase. The result value of $m_\sigma$ depends then on $\lambda_{2,\Lambda}$ and $\Lambda$ itself. This dependence was carefully studied in \cite{gies14} and we shall compare their results with those obtained with the two variants of RGE introduced above. 

 In {\it version A} the explicit RGE is written as a set of first order coupled non-linear equations for the coefficients of the quartic potential. They look like  in the symmetric phase as:
\bea
\partial_t\lambda_{1k}+2\lambda_{1k}&=&
v_d\left[10h_r^2-\frac{2+\lambda_{1k}}{(1+\lambda_{1k})^2}(3\lambda_{2k}+2h_r^2+6\lambda_{1k}h_r^2)+\frac{2h_r^2+3\lambda_{2k}}{1+\lambda_{1k}}\right],
\nonumber\\
\partial_t\lambda_{2k}+(4-d)\lambda_{2k}&=&
v_d\Bigl[-40h_r^4-20\frac{\lambda_{2k}h_r^2(2+\lambda_{1k})}{(1+\lambda_{1k})^2}+\frac{2(2+\lambda_{1k})}{(1+\lambda_{1k})^3}
(3\lambda_{2k}+2h_r^2+6h^2_r\lambda_{1k})^2\nonumber\\
&&~~~~-\frac{2(2h_r^2+3\lambda_{2k})}{(1+\lambda_{1k})^2}
(3\lambda_{2k}+2h_r^2+6h^2_r\lambda_{1k})\Bigr].
\label{sym-lambda-eq}
\eea 
The denominator ($1+\lambda_{1k}$) represents the symmetric phase denominator of the scalar  propagator, while in the symmetric phase the fermion is massless. In the SB phase  the following equations are projected out from the equation valid for a general potential:
\bea
\lambda_{2k}(-\partial_t\kappa_r+(2-d)\kappa_r)&=&
v_d\Biggl[\frac{10h_r^2}{(1+2h_r^2\kappa_r)^2}-\frac{2(1+\kappa_r(\lambda_{2k}+h_r^2))}{(1+2h_r^2\kappa_r)^2(1+2\lambda_{2k}\kappa_r)^2}(3\lambda_{2k}+14h_r^2\lambda_{2k}\kappa_r+2h_r^2)
\nonumber\\
&+&\frac{3\lambda_{2k}+2h_r^2}{(1+2h_r^2\kappa_r)(1+2\lambda_{2k}\kappa_r)}\Biggr].
\label{ssb-minimum}
\eea
\bea
\partial_t\lambda_{2k}+(4-d)\lambda_{2k}&=&v_d\Biggl[-\frac{40h_r^4}{(1+2h_r\kappa_r)^3}-\frac{40h_r^2\lambda_{2k}(1+\kappa_r(\lambda_{2k}+h_r^2))}{(1+2h_r^2\kappa_r)^2(1+2\lambda_{2k}\kappa_r)^2}\nonumber\\
&+&\frac{4(1+\kappa_r(h_r^2+\lambda_{2k}))(2h_r^2+3\lambda_{2k}+14h_r^2\lambda_{2k}\kappa_r)^2}{(1+2h_r^2\kappa_r)^3(1+2\lambda_{2k}\kappa_r)^3}\nonumber\\
&-&\frac{2(2h^2_r+3\lambda_{2k})(2h_r^2+3\lambda_{2k}+14h_r^2\lambda_{2k}\kappa_r)}{(1+2h_r^2\kappa_r)^2(1+2\lambda_{2k}\kappa_r)^2}\Biggr].
\label{ssb-lambda2}
\eea

For {\it version B} the projected equations of the potential are rather similar to those appearing above, just taking into account the subtraction performed in Eq.(\ref{LPA-version-2}). We save space by not showing them explicitly. The equations for the Yukawa coupling are also rather simple:
\bea
\partial_t h_r^2+(4-d)h_r^2&=&4h_r^4v_d\frac{2+\lambda_{1k}}{(1+\lambda_{1k})^2} \qquad {\textrm{(SYM)}},\nonumber\\
\partial_t h_r^2+(4-d)h_r^2&=&8h_r^4v_d\frac{1+\kappa_r(h_r^2+\lambda_{2k})}{(1+2\lambda_{2k}\kappa_r)^2(1+2h_r^2\kappa_r)^2}\qquad {\textrm{(SB)}}. 
\eea

\begin{figure} 
\includegraphics[width=7.5cm]{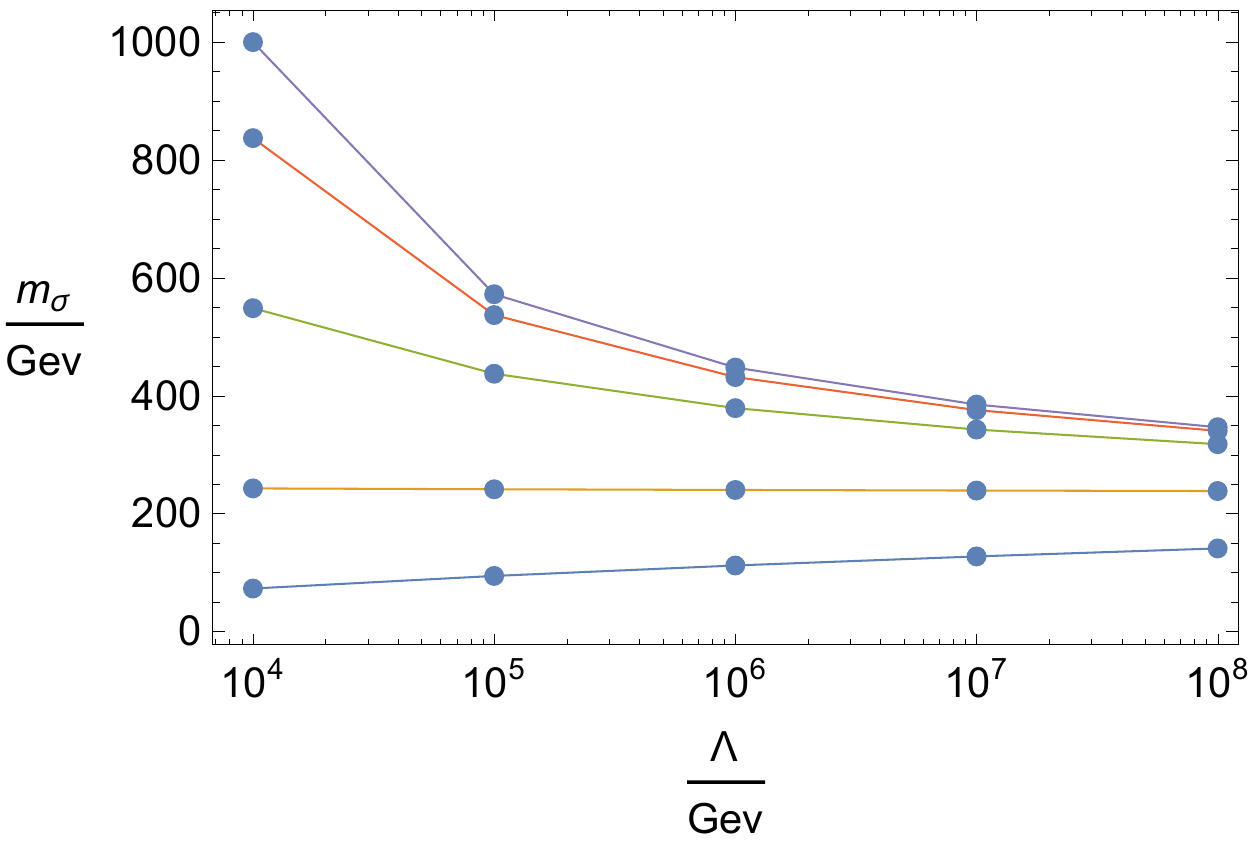}
\includegraphics[width=7.5cm]{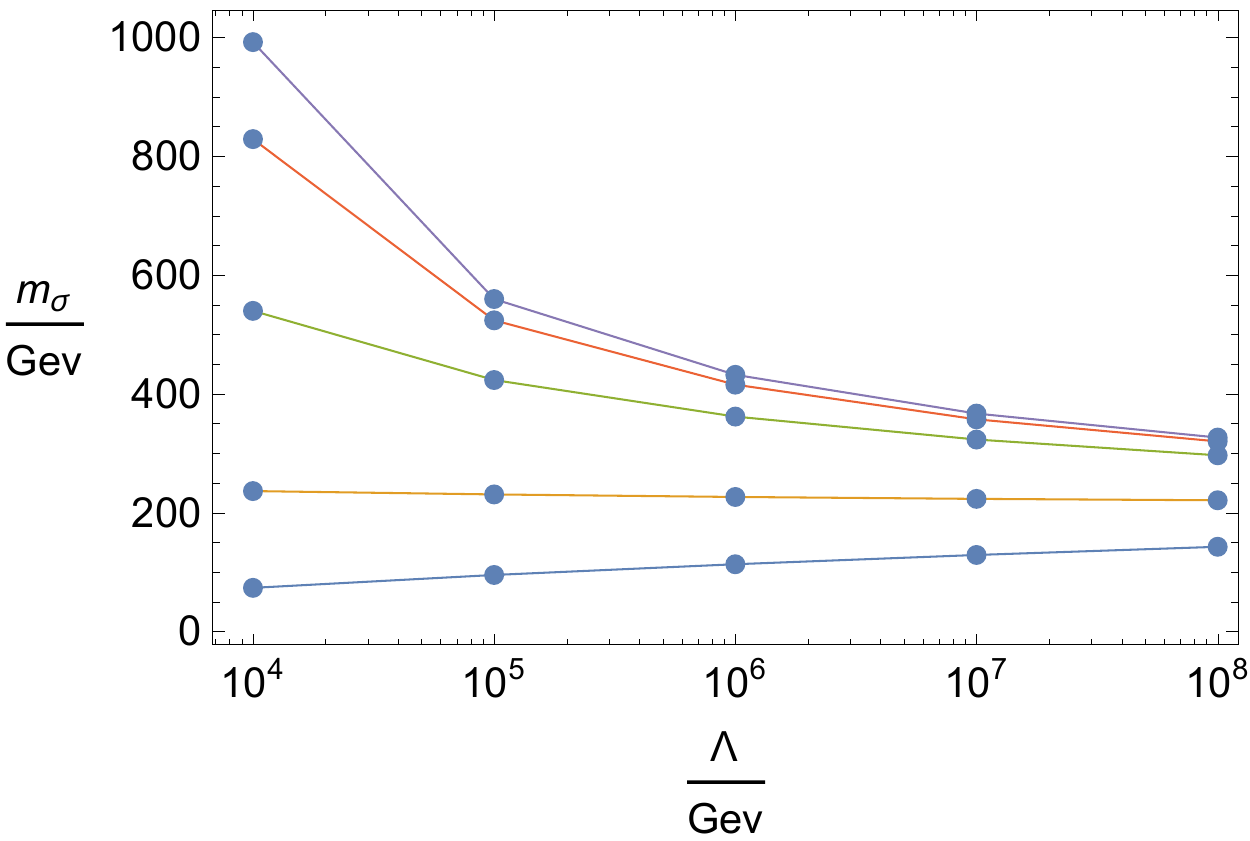}
\caption{The cutoff-dependence of $m_\sigma$ for $\lambda_{2,\Lambda}=0.001,1.,10.,50.,100.$, from below upwards, obtained with {\it version A} (left) and {\it version B} (right) of RGE, respectively. }
\label{Lambda-dep-different-lambda}
\end{figure}
The resulting scalar mass values $m_\sigma$ depend monotonically on $\lambda_{2,\Lambda}$ at any fixed value of $\Lambda$. With increasing $\lambda_{2,\Lambda}$ its value meets a limiting value and apparently a limiting curve $m_\sigma(\lambda_{2,\Lambda}\rightarrow\infty, \Lambda)$ forms. It represents the triviality upper bound on the scalar mass. This is presented in Fig.\ref{Lambda-dep-different-lambda}. The left hand figure corresponds to {\it version A}, the right hand figure to {\it version B}.
The figures are very similar not only qualitatively but also quantitatively, which argues for the robustness of the results against the different backgrounds applied in LPA. 

\begin{figure}
\centerline{ 
\includegraphics[width=10cm]{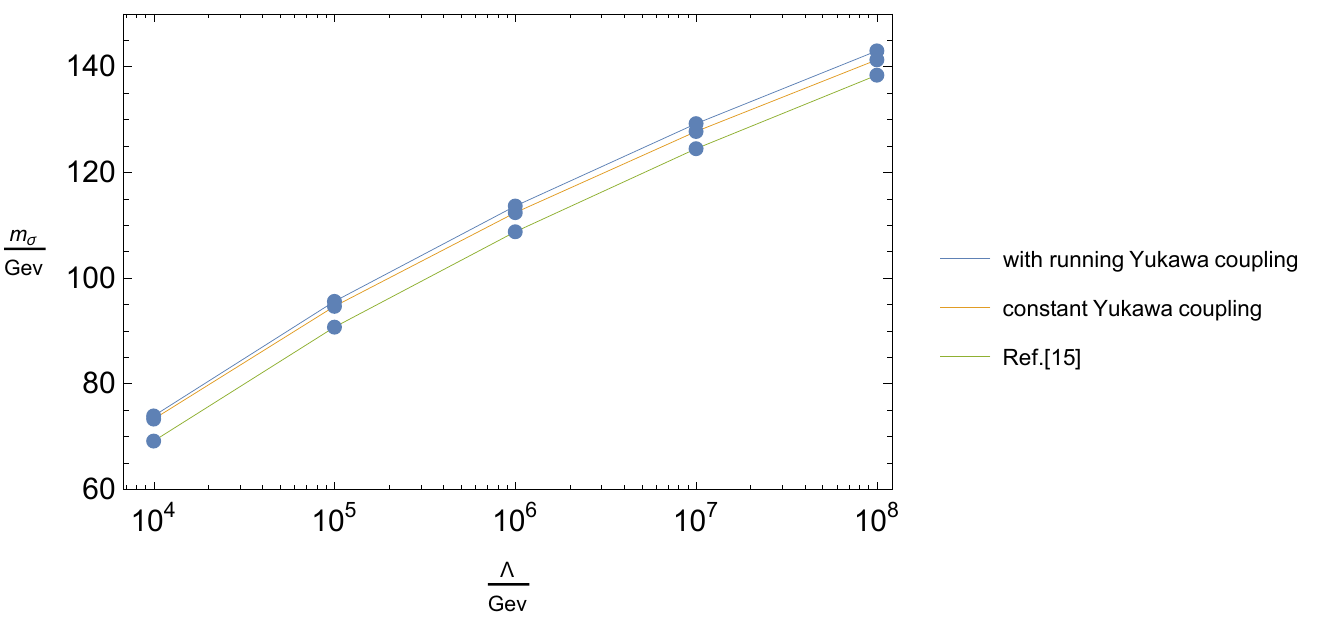}
}
\caption{The cutoff-dependence of $m_\sigma$ for $\lambda_{2,\Lambda}=0.001$  as obtained in three variants of LPA.}
\label{lower-bound-different truncations}
\end{figure}
\begin{figure} 
\includegraphics[width=9cm]{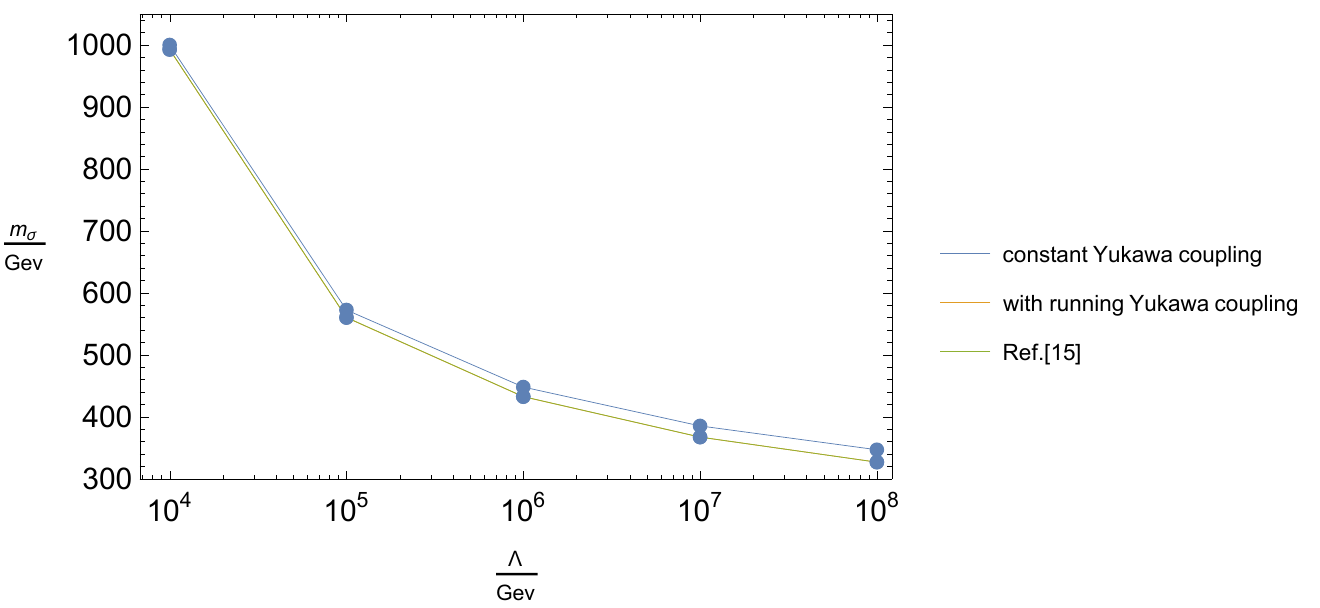}
\includegraphics[width=5.5cm]{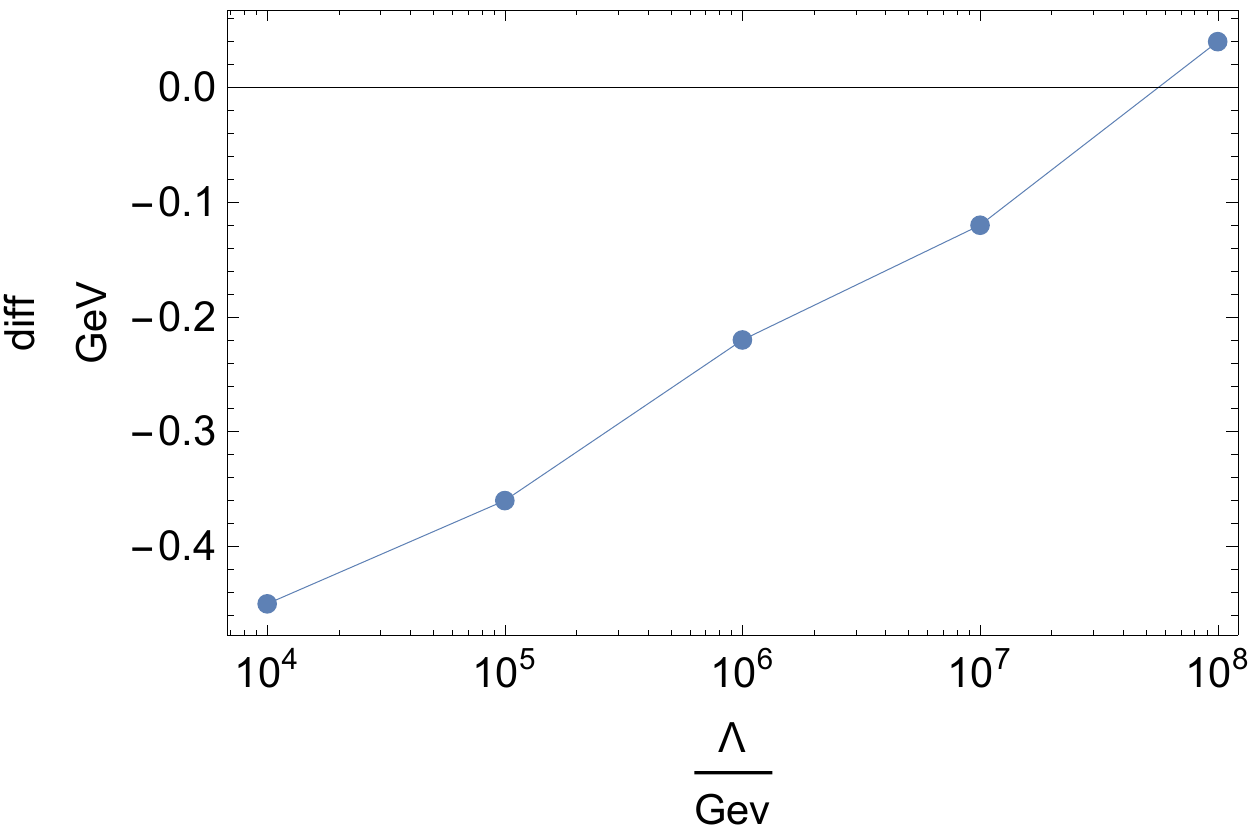}
\caption{The cutoff-dependence of $m_\sigma$ for $\lambda_{2,\Lambda}=100$ (left) as obtained in three variants of LPA. The absolute difference of the estimates arising from our {\it version B} and the RGE used in \cite{gies14} are shown on the right hand figure}
\label{upper-bound-different truncations}
\end{figure}

The stability of the results of \cite{gies14} against including additional contributions due to a non-vanishing $(\bar\psi\psi)_k$ background is within 5\% as one can see from Fig.\ref{lower-bound-different truncations} and Fig.\ref{upper-bound-different truncations}.
Fig.\ref{lower-bound-different truncations} displays the variation of the scalar mass estimate for the smallest value $\lambda_{2,\Lambda}=0.001$ which provides an estimate of the lower $m_\sigma$ bound. One notices that all three estimates lie rather close to each other and follow the same functional dependence with $\Lambda$.

In case of the largest scalar self-coupling $\lambda_{2,\Lambda}=100.$ the points from \cite{gies14} and our {\it version B} where the Yukawa-coupling non-trivially runs are optically indistinguishable, though the $\Lambda$-dependence of the difference shows some systematic tendency. These statements are illustrated in Fig.\ref{upper-bound-different truncations}.

\section{Conclusions}
The non-perturbative mass estimates obtained in the lowest non-trivial truncation of the scalar potential taking into account the effect of a non-trivial pointlike composite fermion background extremely closely coincide with the results of a recent study where only scalar field background was included. This could have been expected with the present truncation, since it can explore only the close neighbourhood of the symmetry breaking minimum, where by Eq.(\ref{effective-field-equation}) the additional diagrams inclu7ded into our treatment in addtion to those of \cite{gies14} give only minor contribution. Certainly, this study should be continued in several aspects, before conclusions drawn from first experience could be generalized. First, higher dimensional operators should be included into the scalar potential and check that they influence the lower bound of $m_\sigma$ as was signalled in \cite{gies14,chu15}. Here the analytic continuation of the Taylor expanded potential into a functional form obeying the RGE dictated asymptotic behavior is of central interest \cite{jakovac15}.  Second, also the effect of the wavefunction renormalisation should be included into the RGE. More complete results on both questions is communicated in the detailed publication \cite{jakovac15b}.

It will be natural to step further to investigating the stability of the Standard Model with the presented method and compare its outcome with some recent results pointing to the metastability of the electroweak vacuum \cite{bergerhoff99,isidori01,branchina15}.

\section*{Acknowledgements}
We thank the organizers of the Gribov-85 Conference, in particular Kolya Nikolaev for the invitation. This research has been supported by the Hungarian Science Fund under the contract OTKA-K104292.

\end{document}